\documentclass{article}
\usepackage[T1]{fontenc} 
\usepackage[utf8]{inputenc} 
\usepackage{ismir,amsmath,url}
\usepackage{graphicx}
\usepackage{booktabs}

\usepackage{color}

\usepackage{array}

\usepackage[firstpage]{draftwatermark}
\definecolor{lightgray}{rgb}{0.9,0.9,0.9}
\definecolor{darkgray}{rgb}{0.4,0.4,0.4}
\SetWatermarkFontSize{12pt}
\SetWatermarkScale{1.1}
\SetWatermarkAngle{90}
\SetWatermarkHorCenter{202mm}
\SetWatermarkVerCenter{170mm}
\SetWatermarkColor{darkgray}
\SetWatermarkText{Late-Breaking / Demo Session Extended Abstract, ISMIR 2024 Conference}

\def\lbd{}	

\title{MIRFLEX: Music Information Retrieval Feature Library for Extraction}

\threeauthors
  {Anuradha Chopra} {\small Singapore University of \\ \small Technology and Design \\ {\tt \small anuradha\_chopra@mymail.sutd.edu.sg}}
  {Abhinaba Roy} {\small Singapore University of \\ \small Technology and Design \\ {\tt \small abhinaba\_roy@sutd.edu.sg}}
  {Dorien Herremans} {\small Singapore University of \\ \small Technology and Design \\ {\tt \small dorien\_herremans@sutd.edu.sg}}

\def\authorname{A. Chopra, A. Roy, and D. Herremans}

\begin{document}

\maketitle
\begin{abstract}
This paper introduces an extendable modular system that compiles a range of music feature extraction models to aid music information retrieval research. The features include musical elements like key, downbeats, and genre, as well as audio characteristics like instrument recognition, vocals/instrumental classification, and vocals gender detection. The integrated models are state-of-the-art or latest open-source. The features can be extracted as latent or post-processed labels, enabling integration into music applications such as generative music, recommendation, and playlist generation. The modular design allows easy integration of newly developed systems, making it a good benchmarking and comparison tool. This versatile toolkit supports the research community in developing innovative solutions by providing concrete musical features.
\end{abstract}
\section{Introduction}\label{sec:introduction}

Music Information Retrieval (MIR) is a complex field focused on computational analysis and processing of musical data, with tasks like similarity estimation, genre classification, and recommendation. While recent advances in machine learning have led to powerful feature extraction methods, the fragmented nature of these tools poses challenges for researchers who must integrate multiple disparate systems. To address this, we present MIRFLEX, a unified feature extraction library designed for MIR research. MIRFLEX offers a diverse set of extractors covering key musical aspects such as key, beats, and genre, using both signal processing and machine learning techniques to generate comprehensive audio representations. 

The primary objectives of this work are threefold: 
\begin{enumerate}
    \item To offer a centralized and easily accessible collection of feature extraction tools, reducing the burden on researchers to implement and integrate disparate feature extraction techniques.  
    \item To provide a comprehensive feature set that captures the multifaceted nature of musical data, enabling researchers to explore a wide range of music-related applications and queries. 
    \item To contribute to the advancement of music information retrieval research by facilitating the rapid prototyping and development of new applications that leverage easily accessible and readily available musical features. 
\end{enumerate}

The proposed feature extraction library is available at \footnote{\url{https://github.com/AMAAI-Lab/megamusicaps}}

%





\section{Integrated Feature Extractors}\label{sec:page_size}

Our feature extractor comprises of the following features. We also detail our selection process for the exact approach for feature extraction:

\subsection{Key Detection}
We consider following approaches for key detection, Inception Key Net and CNNs with Directional Filters. 

\textbf{Inception Key Net} \cite{baumann2021deeper} adapts the Inception V3 architecture and uses the Constant-Q transform to convert the audio to a time-frequency representation. Achieves state-of-the-art performance.

\textbf{CNNs with Directional Filters} \cite{SchreiberM19_CNNKeyTempo_SMC} approach compares shallow, domain-specific architectures with directional filters to deep VGG
-style architectures with square filters, using constant-Q magnitude spectrograms.

\begin{table}
 \begin{center}
 \scriptsize
 \begin{tabular}{llc}
  \toprule
  \textbf{Approach} &\textbf{Dataset}& \textbf{Accuracy(\%)}\\
  \midrule
  InceptionKeyNet \cite{baumann2021deeper}&GiantSteps \cite{knees2015two}& 75.5 \\
  CNNs with Directional Filters \cite{SchreiberM19_CNNKeyTempo_SMC}&GiantSteps \cite{knees2015two}& 67.9 \\
  \bottomrule
 \end{tabular}
\end{center}
\vspace{-0.4cm}
 \caption{Key Detection Extractor candidates.}
 \label{tab:key_detection_extractor_candidates}
\end{table}

In spite of the performance in Table \ref{tab:key_detection_extractor_candidates}, \cite{baumann2021deeper} does not have model weights publicly available. We integrate \cite{SchreiberM19_CNNKeyTempo_SMC} using weights they provided \footnote{\url{https://github.com/hendriks73/key-cnn}}. 

\subsection{Chord Detection}

We consider the following options for chord detection:

\textbf{Bidirectional Transformers}
Uses a self-attention mechanism to focus on regions of chords \cite{park2019bi}. Self-attention is applied forward (on preceding frames) and backward (on succeeding frames) in parallel.  It achieves a weighted chord symbol recall (WCSR) score of 83.9 for the Root chord and 83.1 for the maj-min label.

\textbf{CNN-MCTC with HMM} \cite{weiss2021training} combines Convolutional Neural Networks with  Multi-Class Temporal Classification and Hidden Markov Models. It can be used for both chord recognition and local key estimation.

\textbf{Fully Convolutional Networks with CRF} \cite{korzeniowski2016fully} uses a fully convolutional deep auditory model for feature extraction, followed by CRF for decoding to output the chord sequence.

\textbf{d. Semi-Supervised Jointly Trainable CNN} \cite{wu2020semi} combines CNN with a semi-supervised learning approach , that can jointly train on labelled and unlabelled data to improve chord estimation accuracy.

Overall, we select the Bidirectional Transformer as it achieves the best performance.

\begin{table}
\scriptsize
 \begin{center}
 \begin{tabular}{>{\raggedright\arraybackslash}p{3cm}>{\raggedright\arraybackslash}p{2cm}>{\raggedright\arraybackslash}p{2.1cm}}
  \toprule
  \textbf{Approach} &\textbf{Dataset}& \textbf{Accuracy}\\
  \midrule
  Bidirectional Transformer \cite{park2019bi}&Isophonics Subset + Robbie Williams& 83.9 (WCSR) \\
\midrule
  CNN-MCTC with HMM \cite{weiss2021training}&Schubert Winterreise Dataset& 0.818 (F1 Score) \\
\midrule
  Fully Convolutional Network with CRF \cite{korzeniowski2016fully}&Isophonics Subset& 82.9 (WCSR) \\
\midrule
  Semi-supervised Jointly Trainable CNN \cite{wu2020semi}  &Isophonics + RWC + uspop2002 + McGill Billboard& 82.8 (Accuracy) \\
  \bottomrule
 \end{tabular}
\end{center}
\vspace{-0.4cm}
 \caption{Chord transcription candidates. }
 \label{tab:chord_transcription_extractor_candidates}
\end{table}

\subsection{Down-beat Transcription / Tempo Estimation}

We consider the following approaches for the Downbeat transcription/tempo estimation.

\textbf{CNNs with Directional Filters} \cite{SchreiberM19_CNNKeyTempo_SMC} uses Convolutional Neural Networks, with directional convolutional kernels instead of square ones. 
\textbf{Single-Step Tempo Estimation CNN} \cite{schreiber2018single} frames tempo estimation as a multi-class classification problem, and uses conventional Convolutional Neural Network for the architecture . It can be used for audio clips of 11.9s and therefore is suitable for both local and global tempo estimation. 

\textbf{1D State Space HMM} \cite{heydari2022novel} utilizes a 1D state space and a semi-Markov model for music rhythmic analysis. This approach can reduce the computation cost and provides 30 times speedup in processing. 

\textbf{BeatNet: CRNN and Particle Filtering} \cite{heydari2021beatnet} combines Convolutional-Recurrent Neural Networks with particle filtering for online joint beat, downbeat and meter tracking . It can achieve real-time processing, with high accuracy, albeit computationally expensive. 

Based on its performance, we choose BeatNet.

\begin{table}
\scriptsize
 \begin{center}
 \begin{tabular}{llc}
  \toprule
  \textbf{Approach}  &\textbf{Dataset}& \textbf{Accuracy (\%)} \\
  \midrule
  CNNs with Directional Filters \cite{SchreiberM19_CNNKeyTempo_SMC}  &GiantSteps \cite{knees2015two}& 88.7 \\

  Single Step Tempo Estimation CNN \cite{schreiber2018single}  &GiantSteps \cite{knees2015two}& 86.4 \\

  1D State Space HMM \cite{heydari2022novel}&GTZAN \cite{tzanetakis2002musical}& 76.48 (F-measure) \\

  BeatNet: CRNN and Particle Filtering \cite{heydari2021beatnet}&GTZAN& 80.64 (F-measure) \\
  \bottomrule
 \end{tabular}
\end{center}
\vspace{-0.4cm}
 \caption{Downbeat transcription/tempo estimation.}
 \label{tab:downbeat_trascription_extractor_candidates}
\end{table}

\subsection{Vocals / Instrumental Detection}

The EfficientNet model, trained on Discogs, is used for instrument/vocals and vocals gender detection. The implementation and weights are from the Essentia library \cite{alonso2020tensorflow}. 

We compare this to the vocals \& gender detection implementation available at \footnote{\url{https://github.com/x4nth055/gender-recognition-by-voice}}. Based on results, we choose Discogs-Effnet for vocals / instrumental detection.

\subsection{Instrument, Mood / Theme, Genre Detection}

Not many approaches open-source weights and implementations, we are not able to use any major latest model available. Consequently, the feature extraction system incorporates available weights and techniques implemented in the Essentia library. The Essentia library employs an array of Convolutional Neural Networks as the model architecture, utilizing the Jamendo baseline for the tasks of instrument detection, mood/theme detection, and genre detection.

\begin{table}[h!]
\scriptsize
  \centering
  \begin{tabular}{>{\raggedright\arraybackslash}p{3cm}>{\raggedright\arraybackslash}p{2cm}>{\raggedright\arraybackslash}p{1cm}>{\raggedright\arraybackslash}p{1cm}}
    \toprule
    \textbf{Approach}  &\textbf{Dataset}& \textbf{F1 Score} & \textbf{Number of instruments} \\
    \midrule
    CNN \cite{mahanta2023exploiting}&Philharmonica + UIowa-MIS& 0.98 & 20 \\
        \midrule
    Decoupled CNN per instrument \cite{blaszke2022musical} &Slakh& 0.93 & 4 \\
        \midrule
    SVM classifier w 10-fold cross-validation \cite{racharla2020predominant}&IRMAS& 0.81 & 6 \\
        \midrule
    U-Net Reprogramming \cite{chen2023music}&OpenMIC& 0.81 & 20 \\
        \midrule
    Attention-based Embedding Network \cite{gururani2019attention}&OpenMIC& 0.81 & 20 \\
       \midrule
    Leveraging Hierarchical Structures for few-shot detection \cite{garcia2021leveraging}&MedleyDB& 0.81 & 24 \\
    \bottomrule
  \end{tabular}
  \caption{Instrument detection candidates.}
  \label{tab:instrument_detection_extractor_candidates}
\end{table}

\begin{table}[h!]
    \centering
    \scriptsize
    \begin{tabular}{>{\raggedright\arraybackslash}p{3cm}>{\raggedright\arraybackslash}p{2cm}>{\raggedright\arraybackslash}p{2cm}}
        \toprule
        \textbf{Approach}  &\textbf{Dataset}& \textbf{Accuracy}\\
        \midrule
         Frequency-Aware RF-Regularized CNNs \cite{koutini2019emotion}&MTG-Jamendo \cite{bogdanov2019mtg}& 0.1546 \\
         \midrule
         Noisy student training and Harmonic Pitch Class profiles \cite{tan2021semi}&MTG-Jamendo \cite{bogdanov2019mtg}& 0.1356 \\\bottomrule
    \end{tabular}
    \caption{Mood / Emotion Detection candidates.}
    \label{tab:mood_detection_extractor_candidates}
\end{table}

\section{Call To Contribute}
MIRFLEX is an extendable toolkit freely available on GitHub \footnote{\url{https://github.com/AMAAI-Lab/megamusicaps}}. Moreover, we invite the research community to contribute new feature extractors or extend the existing ones. By providing a common platform for feature extraction, MIRFLEX can facilitate the rapid prototyping and development of new MIR applications, while also enabling researchers to experiment with a diverse set of musical representations. The modular design of MIRFLEX allows for easy integration of new feature extractors, empowering researchers to expand the toolkit's capabilities and push the boundaries of music information retrieval.

We believe that MIRFLEX can serve as a valuable resource for the MIR community, fostering collaboration, reproducibility, and innovation in the field.



\bibliography{ISMIRtemplate}

%
%
%
%
%

\end{document}